\documentclass{emulateapj}
\newcommand{\beq}{\begin{equation}}
\newcommand{\beqa}{\begin{eqnarray}}
\newcommand{\eeq}{\end{equation}}
\newcommand{\eeqa}{\end{eqnarray}}
\newcommand{\simg}{\gtrsim}
\newcommand{\siml}{\lesssim}
\shorttitle{Dispersion Measure from GRB Afterglows}
\shortauthors{Ioka}
\begin{document}
\title{
Cosmic Dispersion Measure from Gamma-Ray Burst Afterglows:
\\
Probing the Reionization History and the Burst Environment
}
\author{
Kunihito Ioka$^{1}$
}
\altaffiltext1{Department of Earth and Space Science,
Osaka University, Toyonaka 560-0043, Japan}

\begin{abstract}
We show a possible way to measure the column density of free electrons 
along the light path, the so-called Dispersion Measure (DM),
from the early $[\sim 415 (\nu/1\ {\rm GHz})^{-2} 
({\rm DM}/10^{5}\ {\rm pc}
\ {\rm cm}^{-3})\ {\rm s}]$
radio afterglows of the gamma-ray bursts.
We find that the proposed Square Kilometer Array
can detect bright radio afterglows 
around the time $\sim 10^{3}(\nu/160\ {\rm MHz})^{-2}$ s
to measure the intergalactic DM
($\simg 6000$ pc cm$^{-3}$ at redshift $z>6$) up to $z\sim 30$,
from which we can determine the reionization history of the universe
and identify the missing warm-hot baryons if many DMs can be measured.
At low $z$, DM in the host galaxy may reach $\sim 10^{5}$ pc cm$^{-3}$
depending on the burst environment, 
which may be probed by the current detectors.
Free-free absorption and diffractive scattering may also affect
the radio emission in a high density.
\end{abstract}


\section{Introduction and summary}
Many observations support
a massive star origin for long-duration Gamma-Ray Bursts (GRBs).
Although the nature of the central engine that accelerates
the GRB jets remains unknown,
the afterglows are successfully fitted by the synchrotron shock model
for probing the jet properties and burst environment
(e.g., M${\acute {\rm e}}$sz${\acute {\rm a}}$ros 2002).

On the other hand, GRBs may be useful for probing high redshift $z$
(e.g., Miralda-Escud$\acute{\rm e}$ 1998;
Barkana \& Loeb 2003; Inoue, Yamazaki, \& Nakamura 2003).
Their high luminosities make them detectable even at $z \sim 100$
(Lamb \& Reichart 2000), 
while the X-ray and infrared afterglows are observable up to $z \sim 30$
(Ciardi \& Loeb 2000; Gou et al. 2003)
and provide their $z$ through the Ly$\alpha$ break (Lamb \& Reichart 2000)
or Fe lines (M${\acute {\rm e}}$sz${\acute {\rm a}}$ros \& Rees 2003),
in contrast to galaxies or quasars which are likely dimmer at higher $z$.
About $25\%$ of all GRBs detected 
by the upcoming $\it Swift$ satellite
are expected to be at $z>5$ (Bromm \& Loeb 2002).
High $z$ GRBs may have already been detected by BATSE
based on the empirical luminosity indicators
(Fenimore \& Ramirez-Ruiz 2000; Norris, Marani, \& Bonnell 2000;
Murakami et al. 2003; Yonetoku et al. 2003),
for which some theoretical explanations exist (e.g., Ioka \& Nakamura 2001a).
The first generation stars could be very massive
(Abel, Bryan, \& Norman 2002; Bromm, Coppi, \& Larson 2002;
Omukai \& Palla 2003),
so that they may end as brighter GRBs.

In this Letter, we show that free electrons along the light path
cause distortion in the spectrum and light curve of an early radio afterglow,
from which we can measure the column density of the free electrons,
the so-called Dispersion Measure (DM).\footnote{
Inoue (2003) has also independently carried out a similar study.}
This is because, in a plasma with an electron density $n_{e}$,
an electromagnetic wave with frequency $\nu (\gg \nu_{p})$
is delayed relative to in a vacuum by a time
\beqa
\Delta t\simeq \int \frac{dl}{c} \frac{\nu_{p}^{2}}{2\nu^{2}}
=415 \left(\frac{\nu}{1\ {\rm GHz}}\right)^{-2}
\left(\frac{\rm DM}{10^{5}\ {\rm pc}\ {\rm cm}^{-3}}\right)\ {\rm s},
\label{eq:dt}
\eeqa
where $\nu_{p}=(n_{e} e^2/\pi m_{e})^{1/2}=8.98 \times 10^{3} n_{e}^{1/2}$ Hz
is the plasma frequency and ${\rm DM}=\int n_{e} dl$
(Rybicki \& Lightman 1979).
DMs of pulsars are well studied by using the arrival time 
of pulses at two or more frequencies 
(Taylor, Manchester, \& Lyne 1993).
Our method to measure DMs of GRBs is somewhat different
since we do not assume simultaneously emitted pulses
but the afterglow model.
We can determine DM only from 
a single-band light curve around the time $t\sim \Delta t$
(see \S \ref{sec:after}).
The former method was suggested by Ginzburg (1973)
and Palmer (1993) before the discovery of the afterglows.

At $z>6$, DM$_{\rm IGP}$ due to the intergalactic plasma (IGP)
will be $\simg 6000$ pc cm$^{-3}$ 
[correspondingly $\Delta t \simg 10^{3}(\nu/160\ {\rm MHz})^{-2}$ s],
and probably dominate
DM$_{\rm G}$ due to the Galactic plasma
and DM$_{\rm host}$ due to the plasma in the host galaxy
(see \S \ref{sec:dm}).
DM$_{\rm IGP}$ as a function of $z$ varies depending on the 
reionization history of the universe since recombined electrons provide no DM
(see \S \ref{sec:reion}).
Thus, from DMs of GRBs at various $z$ and directions,
we can determine the reionization history,
and possibly even map the topology of the ionized bubbles.
The reionization history is now actively investigated but
remains unclear (Miralda-Escud$\acute{\rm e}$ 2003).
The analysis of the Ly$\alpha$ spectra in 
the highest $z$ quasars suggests that the reionization ends
at $z\sim 6$ (Fan et al. 2002),
while the WMAP polarization data imply a much higher reionization
redshift $z\sim 17\pm 5$ (Kogut et al. 2003; Spergel et al. 2003).
In \S \ref{sec:reion},
we show that the proposed Square Kilometer Array (SKA)
can detect bright afterglows at $t\sim \Delta t$
to measure DM$_{\rm IGP}$ up to $z\sim 30$.
In addition, observations of DM$_{\rm IGP}$ could also identify
the 'missing' baryons that have not been detected at $z\siml 1$
(see \S \ref{sec:dm}).

At low $z$, DM$_{\rm host}$ in the host galaxy 
may reach $\sim 10^{5}$ pc cm$^{-3}$ and 
dominate DM$_{\rm G}$ and DM$_{\rm IGP}$
depending on the as-yet-unknown burst environment
(see \S \ref{sec:dm}).
Since we may target low $z$ GRBs, 
even current detectors could detect afterglows at $t\sim \Delta t$
to constrain DM$_{\rm host}$ and the GRB environment.
Throughout we adopt a $\Lambda$CDM cosmology
with $(\Omega_{m}, \Omega_{\Lambda}, \Omega_{b}, h, \sigma_{8})
=(0.27, 0.73, 0.044, 0.71, 0.84)$ (Spergel et al. 2003).

\section{Expected Dispersion Measure (DM)}\label{sec:dm}

{\it Galactic DM$_{G}$}:
The distribution of free electrons in our Galaxy is relatively well known
from pulsar DMs (Taylor \& Cordes 1993).
DM$_{\rm G}$ has its maximum (minimum)
in the direction parallel (perpendicular) to the Galactic plane,
ranging from 
\beqa
{\rm DM}_{\rm G}^{\rm min}\sim 30\ {\rm pc}\ {\rm cm}^{-3},
\nonumber
\eeqa
to ${\rm DM}_{\rm G}^{\rm max}\sim 10^{3}$ pc cm$^{-3}$ 
(Taylor et al. 1993;
Nordgren, Cordes, \& Terzian 1992).

\begin{figure}
\plotone{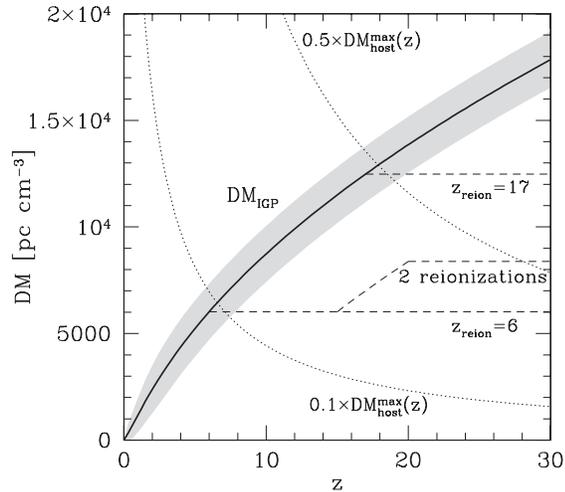}
\caption{\label{fig:dm}
The Dispersion Measure (DM) due to the intergalactic plasma, DM$_{\rm IGP}$,
is plotted as a function of redshift $z$ by the solid line.
The dispersion of DM$_{\rm IGP}$ due to density fluctuations 
is shown by the shaded region.
$0.5\times {\rm DM}_{\rm host}^{\rm max}(z)$
and $0.1\times {\rm DM}_{\rm host}^{\rm max}(z)$
are also shown by the dotted lines where
DM$_{\rm host}^{\rm max}(z)={\rm DM}_{\rm host}^{\rm max}/(1+z)$
is the maximum DM due to the plasma in the host galaxy in
equation (\ref{eq:dmmax}).
The dashed lines are DM$_{\rm IGP}$ 
for a sudden reionization at $z_{\rm reion}=6$,
$z_{\rm reion}=17$ and the case when the reionization occurred twice
at $z_{\rm reion}=6, 20$ with a full recombination at $z=15$.
}
\end{figure}

{\it Intergalactic DM$_{IGP}$}:
We first estimate DM$_{\rm IGP}$ assuming that all baryons are fully ionized
and homogeneously distributed, so that the free electron density evolves as 
$n_{e}=(3H_{0}^{2}\Omega_{b}/8\pi G m_{p})$ $(1+z)^{3}$.
Then the observed dispersion delay at an observed frequency $\nu$
for a source at $z$ is given by
$\Delta t=\int_{0}^{z} dz |dt/dz| (1+z)
\nu_{p}^{2}/2 \left[\nu (1+z)\right]^{2}
=415 \left(\nu/1\ {\rm GHz}\right)^{-2}$
$\left({\rm DM}_{\rm IGP}/10^{5}\ {\rm pc}\ {\rm cm}^{-3}\right)$ s, 
where $|dt/dz|^{-1}=(1+z) H_{0}
\left[\Omega_{m} (1+z)^{3}+\Omega_{\Lambda}\right]^{1/2}$
and 
\beqa
{\rm DM}_{\rm IGP}=\frac{3 c H_0 \Omega_{b}}{8 \pi G m_{p}}
\int_{0}^{z}
\frac{(1+z) dz}
{\left[\Omega_{m} (1+z)^{3}+\Omega_{\Lambda}\right]^{1/2}}.
\eeqa
From Figure \ref{fig:dm}, 
DM$_{\rm IGP}>$DM$_{\rm G}^{\rm min}$ at $z>0.03$.

More than half of all baryons
have not been detected at $z\siml 1$
(Fukugita, Hogan, \& Peebles 1998).
Most of such 'missing' baryons may reside in a warm-hot IGP,
which is difficult to observe because of its high temperature
($10^{5}$-$10^{7}$ K)
and low density (moderate overdensity $10$-$40$)
(Cen \& Ostriker 1999; Dav$\acute{\rm e}$ et al. 2001).
If so, an appreciable fraction of DM$_{\rm IGP}$ ($\simg 500$ pc cm$^{-3}$)
comes from the missing baryons at $0 \le z < 1$ (see Figure \ref{fig:dm}).
Thus observations of DM$_{\rm IGP}$ could be a direct detection
of the missing baryons.

{\it DM$_{host}$ in host galaxy}:
A host galaxy has been found in most GRBs 
(Bloom, Kulkarni, \& Djorgovski 2002).
If the host galaxies are similar to ours, they have
DM$_{\rm host} \sim {\rm DM}_{\rm G}$.
At high $z$, the maximum and minimum DM$_{\rm host}$
may evolve as $\sim$DM$_{\rm G}^{\rm max}
(M/10^{12} M_{\odot}) (1+z)^{3}$
and $\sim$DM$_{\rm G}^{\rm min}
(M/10^{12} M_{\odot})^{1/3} (1+z)^{2}$, respectively, for
a host galaxy with the halo mass $M$ 
based on the hierarchical galaxy formation
(Ciardi \& Loeb 2000).
Even at low $z$, DM$_{\rm host}$ may be high
if GRBs arise from the star forming regions,
as expected in GRBs resulting from massive stellar collapses.
The hydrogen column density is as high as $10^{22}$-$10^{23}$ cm$^{-2}$
in giant molecular clouds (Galama \& Wijers 2001),
and the prompt and afterglow emission with energy $E_{\rm ion}$ erg 
can ionize the ambient density $n_{\rm host}$ cm$^{-3}$ out to a distance of 
$d \sim 160 n_{\rm host}^{-1/3} E_{{\rm ion},52}^{1/3}$ pc 
(Perna \& Loeb 1998),
where the convention $Q=10^{x} Q_{x}$ is used.
Thus all the ambient hydrogen may be ionized depending on its density.
In this case,
${\rm DM}_{\rm host} \sim 10^{3}$-$10^{5}\ {\rm pc}\ {\rm cm}^{-3}$,
which may be time-dependent (Perna \& Loeb 1998).
A high DM is also expected in the 'supranova' model (Vietri \& Stella 1998),
in which a supernova (SN) occurs weeks before the GRB.
A SN remnant shell with mass $M$ g and velocity $v$ cm s$^{-1}$ reaches 
at $R \sim 10^{16} v_{9} t_{7}$ cm in $t$ s,
which provides
${\rm DM}_{\rm host} \sim {M}/{4\pi R^{2} m_{p}}
\sim 1.5 \times 10^{5} v_{9}^{-2} t_{7}^{-2} M_{33}
\ {\rm pc}\ {\rm cm}^{-3}$.

Anyway, 
the maximum DM$_{\rm host}$ allowed by the present observations
is about the inverse of the Thomson cross section $\sigma_{T}$,
\beqa
{\rm DM}_{\rm host}^{\rm max}
\sim \sigma_{T}^{-1}
\sim 4.87 \times 10^{5}
\ {\rm pc}\ {\rm cm}^{-3},
\label{eq:dmmax}
\eeqa
for GRBs and afterglows to be observed without scattering.
Correspondingly,
$\Delta t_{\rm host} \sim 2022 (\nu_{\rm host}/1\ {\rm GHz})^{-2}$ s
from equation (\ref{eq:dt}).
At high $z$, the observed dispersion delay and frequency are 
$\Delta t=(1+z) \Delta t_{\rm host}$ and 
$\nu=\nu_{\rm host}/(1+z)$,
respectively.
Thus it is convenient to use 
${\rm DM}_{\rm host}(z)={\rm DM}_{\rm host}/(1+z)$
to compare it with DM$_{\rm G}$ and DM$_{\rm IGP}$.
From Figure \ref{fig:dm},
DM$_{\rm host}(z)$ declines with increasing $z$
while DM$_{\rm IGP}$ grows.
Therefore, GRBs at high (low) $z$ are suitable 
for measuring DM$_{\rm IGP}$ (DM$_{\rm host}$).

\section{DM from dispersed afterglows}\label{sec:after}
We now study the effects of the DM on the afterglow emission.
For illustrative purposes we fix DM demanding
the observed dispersion delay of $\Delta t=10^{3} (\nu/1\ {\rm GHz})^{-2}$ s 
in this section.
We consider the most simple but standard afterglow model 
(Sari, Piran, \& Narayan 1998),
which reproduces the observations very well.

The afterglow spectrum for the forward shock
is well approximated by four power law segments with breaks
at cooling frequency $\nu_{c}$, typical frequency $\nu_{m}$
and self-absorption frequency $\nu_{a}$.
In the fast cooling case, $(\nu_{a}<)\nu_{c}<\nu_{m}$,
the spectrum is $F_{\nu}\propto \nu^{2}, \nu^{1/3}, \nu^{-1/2}, \nu^{-p/2}$ 
from low to high frequencies,
and has a maximum flux $F_{\nu,\rm max}$ at $\nu=\nu_{c}$.
In the slow cooling, $(\nu_{a}<)\nu_{m}<\nu_{c}$,
it is $F_{\nu}\propto \nu^{2}, \nu^{1/3}, \nu^{-(p-1)/2}, \nu^{-p/2}$, 
and peaks at $\nu=\nu_{m}$.
Assuming an adiabatic shock,
we have the observed break frequencies and maximum flux as
\beqa
\nu_{c}&=&2.5 \times 10^{13} \epsilon_{B}^{-3/2} E_{52}^{-1/2}
n^{-1} t_{3}^{-1/2} (1+z)^{-1/2}\ {\rm Hz},
\nonumber\\
\nu_{m}&=&4.2 \times 10^{18} \epsilon_{B}^{1/2} \epsilon_{e}^{2} g^{2} 
E_{52}^{1/2} t_{3}^{-3/2} (1+z)^{1/2}\ {\rm Hz},
\nonumber\\
F_{\nu,\rm max}&=&1.1 \times 10^{5} \epsilon_{B}^{1/2} E_{52} n^{1/2}
D_{28}^{-2} (1+z)\ \mu {\rm Jy},
\nonumber
\eeqa
where $E$ erg is the isotropic equivalent shock energy,
$n$ cm$^{-3}$ is the constant surrounding density,
$\epsilon_{e}$ ($\epsilon_{B}$) is the electron (magnetic) energy fraction,
$t$ s is the observer time,
$D$ cm is the luminosity distance and $g=(p-2)/(p-1)$.
At $\nu<\nu_{a}$, synchrotron self-absorption limits the flux below
the blackbody emission with the electron temperature
(e.g., Sari \& Piran 1999; Kobayashi \& Sari 2000).
This is given by
\beqa
F_{\nu,\rm BB}=2\pi \nu^{2} \gamma \gamma_{e} m_{e}
\left({R_{\perp}}/{D}\right)^{2} (1+z),
\nonumber
\eeqa
where $\nu$ is the observed frequency,
$\gamma$
is the Lorentz factor of the shocked fluid,
$R_{\perp}=4 \gamma c t$
is the observed size of the afterglow
and $\gamma_{e}$ is the typical Lorentz factor of the electrons
emitting at $\nu$.
A reverse shock 
is also formed,
but its temperature is lower because of its higher density,
making its emission dimmer at $\nu<\nu_{a}$.

\begin{figure}
\plotone{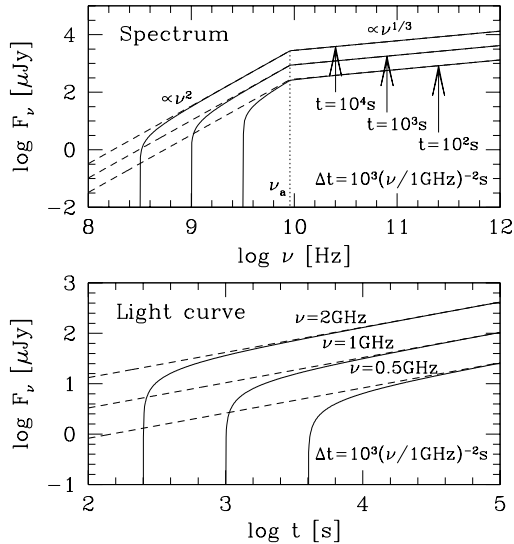}
\caption{\label{fig:spec}
The spectra ({\it upper panel})
and light curves ({\it lower panel})
of afterglows are shown 
with and without the dispersion delay 
$\Delta t=10^{3}(\nu/1\ {\rm GHz})^{-2}$ s
by the solid and dashed lines, respectively.
We adopt $E=10^{52}$ erg, $n=1$ cm$^{-3}$, $\epsilon_{e}=0.1$,
$\epsilon_{B}=0.01$, $p=2.2$ and z=0.5.
For these parameters, the afterglows in this figure
are in the slow cooling regime.
}
\end{figure}

Once we calculate the spectrum as a function of time $F_{\nu} (t)$,
we can obtain a dispersed afterglow as $F_{\nu}[t-\Delta t(\nu)]$
where $\Delta t(\nu)$ is the dispersion delay.
From Figure \ref{fig:spec}, the spectrum has a cutoff at 
$\nu\sim 1 (t/\Delta t_{\rm GHz})^{-1/2}$ GHz
when $\Delta t=\Delta t_{\rm GHz} (\nu/1\ {\rm GHz})^{-2}$
where we are using $\Delta t_{\rm GHz}=10^{3}$ s.
The light curve deviates from the power law at $t\sim \Delta t$.
Measurements of these features yield $\Delta t$ and hence
DM with equation (\ref{eq:dt}).
In practice, DM may be treated as an additional fitting parameter.
Only a single-band light curve will suffice to measure DM,
but a multiband observation will be useful to exclude
other interpretations, such as variable light curves
seen in GRB 021004 and GRB 030329 (e.g., Uemura et al. 2003)
and microlensing (Loeb \& Perna 1998; Ioka \& Nakamura 2001b),
by using the frequency dependence $\Delta t \propto \nu^{-2}$.

\section{Prospects for probing the reionization}\label{sec:reion}

DM$_{\rm IGP}$ depends on the reionization history
(see Figure \ref{fig:dm}).
DM$_{\rm IGP}$ is a constant $\sim 6000$ pc cm$^{-3}$ at $z>z_{\rm reion}$
if a sudden reionization occurred at $z_{\rm reion}=6$,
while DM$_{\rm IGP}\sim 12000$ pc cm$^{-3}$ if $z_{\rm reion}=17$.
We also show the case when the reionization occurred twice
at $z_{\rm reion}=6, 20$ with a full recombination at $z=15$
(Cen 2003a,b; Wyithe \& Loeb 2003).
Contrary to the Ly$\alpha$ absorption (Inoue et al. 2003),
DM can determine the reionization history 
even if the neutral fraction is $\simg 10^{-5}$
in the first reionized era.
If DM$_{\rm host}(z)>$DM$_{\rm IGP}$, the mean DM of the GRBs 
will be above the line of DM$_{\rm IGP}$ in Figure \ref{fig:dm}.

\begin{figure}
\plotone{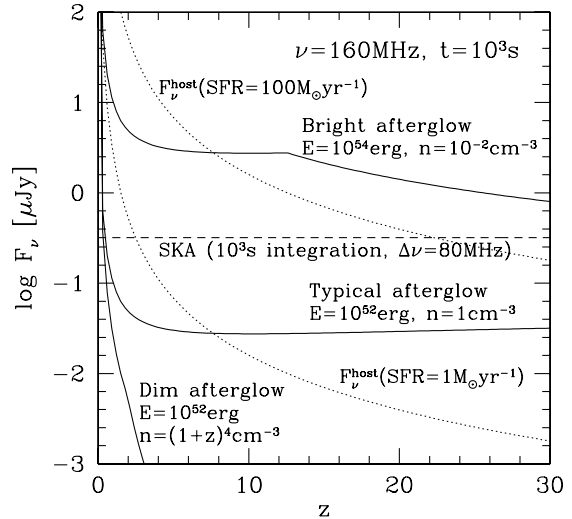}
\caption{\label{fig:zflux}
The afterglow flux at a frequency $\nu=160$ MHz
and an observer time $t=10^{3}$ s
is shown as a function of redshift $z$ by the solid lines.
We adopt $\epsilon_{e}=0.1$, $\epsilon_{B}=0.01$ and $p=2.2$
with $E=10^{52}$, $10^{54}$, $10^{52}$ erg
and $n=1$, $10^{-2}$, $(1+z)^{4}$ cm$^{-3}$ 
for typical, bright, dim afterglows, respectively.
The rms sensitivity of the Square Kilometer Array
for an integration time $10^{3}$ s
and a band width $\Delta \nu\sim 0.5\nu$
is plotted by the dashed line.
The expected flux of the host galaxy 
$F_{\nu}^{\rm host}\sim 25 \left[{(1+z)\nu}/{1\ {\rm GHz}}\right]^{-0.75}
({\rm SFR}/{1\ M_{\odot}\ {\rm yr}^{-1}})
\left({D}/{1\ {\rm Mpc}}\right)^{-2}
(1+z)\ {\rm Jy}$
is shown by the dotted lines
where SFR is the star formation rate (Yun \& Carilli 2002).
}
\end{figure}

Can we measure DMs of high $z$ GRBs ?
We should start a radio follow-up
within $\sim \Delta t \sim 972 \left(\nu/160\ {\rm MHz}\right)^{-2}
\left({\rm DM}/6000\ {\rm pc}\ {\rm cm}^{-3}\right)$ s,
since DM distorts the light curve at $t\sim \Delta t$
(see Figure \ref{fig:spec}).
This may be possible since $\it Swift$
will send a $1$-$4$ arcmin GRB position to the ground 
within $\sim 1$ min (see http://swift.gsfc.nasa.gov/).
The dispersion delay $\Delta t$ is longer at lower frequencies,
but the flux decays as $F_{\nu}\propto \nu^{2}$ at $\nu<\nu_{a}$.
Here we adopt $\nu=160$ MHz
assuming 
a response time of $t\sim 10^{3}$ s.
In Figure \ref{fig:zflux},
the afterglow flux at $\nu=160$ MHz and $t=10^{3}$ s is plotted
as a function of $z$.
For the typical parameters,
$E=10^{52}$ erg, $n=1$ cm$^{-3}$, $\epsilon_{e}=0.1$, $\epsilon_{B}=0.01$ 
and $p=2.2$,
the flux is too dim to be detected
even by SKA, whose rms sensitivity is
$\sim 0.3 \left({\Delta \nu}/{80\ {\rm MHz}}\right)^{-1/2}
\tau_{3}^{-1/2}\ \mu{\rm Jy}$
for a band width $\Delta \nu\sim 0.5\nu$,
an integration time $\tau=10^{3}$ s
and $A_{\rm eff}/T_{\rm sys}=2\times 10^{8}$ cm$^{2}$ K$^{-1}$
(see http://www.astron.nl/skai/science/).
However some GRBs have larger energy $E\sim 10^{54}$ erg
and lower density $n \sim 10^{-2}$ cm$^{-3}$
(Panaitescu \& Kumar 2002).
From Figure \ref{fig:zflux},
such GRBs
can be detected as $\sim 7 \sigma$ events even at $z=15$ by SKA,
and dominate the host galaxy emission
even if ${\rm SFR} \sim 100 M_{\odot}$ yr$^{-1}$ at $z\simg 8$.
Note that the density around the first stars 
could be $10^{-2} \siml n \siml 1$ cm$^{-3}$ 
because of strong radiation pressure from the central massive star
(Gou et al. 2003).
On the contrary, the density could evolve as
$n \propto (1+z)^{4}$ for a fixed host galaxy mass based on the hierarchical
galaxy formation (Ciardi \& Loeb 2000).
If $n\sim (1+z)^{4}$ cm$^{-3}$, 
we will never detect radio afterglows at high $z$.

\section{Discussions}\label{sec:dis}

Free-free absorption (FF) may be important 
when the ambient density $n_{\rm host}$ cm$^{-3}$ is high.
The host galaxy is optically thick to FF 
at $\nu \siml 2 ({\rm DM}_{\rm host}/10^{5}\ {\rm pc}\ {\rm cm}^{-3})^{1/2} 
T_{4}^{-3/4} n_{{\rm host},3}^{1/2} (1+z)^{-1}$ GHz
where $T$ is the plasma temperature in K 
(Rybicki \& Lightman 1979).
In this case we cannot measure DM but can probe the burst environment.
We may neglect FF in IGP at
$\nu \simg 1.6 h^{3/2} \Omega_{b,-2} \Omega_{m,-1}^{-1/4} T_{4}^{-3/4}
(1+z)^{5/4}$ kHz where $z\gg 1$ (Rees 1978).

Large-scale density fluctuations, as well as the accumulation of gas 
in discrete galaxies, 
cause a dispersion in DM$_{\rm IGP}$.
For example, if the average number of intervening gas clumps 
with a DM of $\Sigma$ is $N$,
the dispersion $\langle (\Delta {\rm DM})^{2}\rangle$ is 
$\sim N \Sigma^{2}$ if we assume a Poisson statistic.
We may roughly estimate $\langle (\Delta {\rm DM})^{2}\rangle$ 
using the Press-Schechter theory.
Kitayama \& Suto (1996) derived the comoving number density of halos 
that form with mass $M \sim M+dM$ at time $z_{f}\sim z_{f}+dz_{f}$
and are observed at $z$, $F(M,z_{f};z) dM dz_{f}$.
If the halo mass is $M \siml 10^{12} M_{\odot}$,
the inside gas can cool to have a radius $r_{d}\sim 0.035 r_{\rm vir}$ 
(e.g., Ciardi \& Loeb 2000), otherwise $r_{d}\sim r_{\rm vir}$,
where $r_{\rm vir}(M,z_{f})$ is the virial radius of the halo.
Then, DM of one halo is about $\Sigma(M,z_{f};z)=
M \Omega_{b} /2\pi r_{d}^{2} m_{p} \Omega_{m} (1+z)$,
and the dispersion of DM due to halos in a logarithmic mass interval
is about $\langle [\Delta {\rm DM}(M)]^{2}\rangle \sim
\int c dt \int dz_{f} \pi r_{d}^{2} M F (1+z)^{3} \Sigma^2$.
We calculate $\max \langle [\Delta {\rm DM}(M)]^{2}\rangle$
and show it as the shaded region in Figure \ref{fig:dm}.
It takes the maximum when $M\sim 10^{12} M_{\odot}$ and
is about $\langle [\Delta {\rm DM}(10^{12} M_{\odot})]^{2}\rangle/
{\rm DM}^{2}\sim (68\%)^{2}$ at $z=1$,
though the mean number of intervening halos is $\sim 10^{-3}$.
Unvirialized objects may have less contributions
since $\langle [\Delta {\rm DM}(2\times 10^{14} M_{\odot})]^{2}\rangle/
{\rm DM}^{2} \sim 8 h^{-1}$ Mpc/$6.6$ Gpc $\sim (4\%)^{2}$ at $z=1$
for density fluctuations of radius $8 h^{-1}$ Mpc.

At $\nu \siml \nu_{s} = 10$ GHz,
diffractive and refractive scintillation due to the Galactic plasma take place
(Goodman 1997; Walker 1998).
Although the modulation index (rms fractional flux variation)
is unity $m_{s}=1$ for diffractive scintillation,
the decorrelation bandwidth is very narrow
$\Delta \nu_{s}/\nu = (\nu/\nu_{s})^{17/5}\sim 4\times 10^{-4} \nu_{9}^{17/5}$
at low frequencies.
Then $m_{s}$ will be reduced to $\sim \sqrt{\Delta \nu_{s}/\nu}
\sim 2 \nu_{9}^{17/10} \%$ for a broadband observation.
It will be also reduced when the integration time is longer 
than the scintillation one $\sim 700 \nu_{9}^{6/5}$ s.
For refractive scintillation,
$m_{s}=(\nu/\nu_{s})^{17/30}\sim 27 \nu_{9}^{17/30}\%$.
Thus Galactic scintillation may
be quenched at $\nu \siml 1$ GHz with $\Delta \nu\sim \nu$.
On the other hand, scattering in the host galaxy may cause
temporal broadening of the early radio afterglows by a time
$\sim 1/2\pi \Delta \nu_{s} \sim 10 \nu_{9}^{-22/5} 
({\rm DM}_{\rm host}/10^{5}\ {\rm pc}\ {\rm cm}^{-3})^{6/5}
n_{{\rm host},3}^{6/5} d_{1} (1+z)^{-17/5}$ s,
if the scattering measure scales as $\propto l n_{e}^{2}$
(Goodman 1997), where $d$ pc is the size of the ionized region.

Some GRBs may have the wind environment (Chevalier \& Li 2000).
Inverse Compton may prolong the fast cooling regime 
(Sari \& Esin 2001), in which an additional spectral segment
$F_{\nu}\propto \nu^{11/8}$ may appear (Granot, Piran, \& Sari 2000).
In addition, photons accumulating around the shock front
may make the cooling 'ultrafast' (Ioka 2003).
These are interesting future problems.

\acknowledgments
We thank F.~Takahara, J.~Yokoyama, K.~Omukai and M.~Matsumiya 
for useful discussions.
This work was supported in part by the Monbukagaku-sho Grant-in-Aid
No.~00660.




%
%

%
%




\end{document}